Original Paper

# Deep Learning Approach to Forecasting COVID-19 Cases in Residential Buildings of Hong Kong: A Multi-source Data Analysis of the Differential Roles of Environment and Sociodemographics during the Emergence and Resurgence of the Pandemic

E. Leung[1#], J. Guan[1#], KO. Kwok[1], CT. Hung[1], CC. Ching[1], KC. Chong[1], CHK. Yam[1], T. Sun[1], WH. Tsang[2], EK. Yeoh[1], A. Lee[1*]

1 JC School of Public Health and Primary Care, The Chinese University of Hong Kong
2 Department of Rehabilitation Science, Hong Kong Polytechnic University
# first author
* corresponding author

## Abstract

**Background:** In isolated studies, a building's internal and external built environment and its residents' sociodemographic profiles were separately linked to COVID-19 infections. Yet, their marginal and relative contributions had never been quantified or examined systematically as intricately related constituting elements of the residents' broader socioecology. Most importantly, the epidemic resurgence of COVID-19 in Hong Kong (HK; between 2021-12-24 and 2022-07-23) differs from previous waves of outbreaks regarding virology, clinical characteristics, and the associated public health and social measures.
**Objective:** To investigate the complex association among factors and quantify their contributions to the early outbreaks and epidemic resurgence of COVID-19.
**Methods:** We aligned the analytic model's architecture with the hierarchical structure of the resident's socioecology. Specifically, we applied a multi-headed hierarchical convolutional neural network to structure the vast array of hierarchically related predictive features representing buildings' internal and external built environments and residents' sociodemographic profiles to model COVID-19 cases accumulated in buildings across three adjacent districts in HK, both before and during HK's epidemic resurgence. Most notably, to ensure the validity of the model's predictability beyond the model-training period, a forward chaining validation was performed to examine the model's performance in forecasting COVID-19 cases over the 3-, 7-, and 14-day horizons during the two months subsequent to when the model for COVID-19 resurgence was built, to align with the forecasting needs in an evolving pandemic.
**Results:** Different sets of factors were found to be linked to the earlier waves of COVID-19 outbreaks compared to the epidemic resurgence of the pandemic.

Sociodemographic factors such as work hours, monthly household income, employment types, and the number of non-working adults or children in household populations were of high importance to the studied buildings' COVID-19 case counts during the early waves of COVID-19. Factors constituting one's internal built environment, such as the number of distinct households in the buildings, the number of distinct households per floor, and the number of floors, corridors, and lifts, had the greatest unique contributions to the building-level COVID-19 case counts during epidemic resurgence.
**Conclusions:** Adding to the series of lessons learned the literature has derived from previous phases of COVID-19, our findings that specific internal built environment elements and sociodemographic factors of residential buildings differentially subject residents to COVID-19 infection can inform targeted surveillance and public health and social measures for mitigating the impact of the potential resurgence of COVID-19 in high-risk geographical areas. We provided an approach on how stable elements in one's socioecology can add significant value to forecasting the resurgence of COVID-19 beyond forecasting with historical case count alone, making the development of long-term policy as relevant to reactionary public health and social measures to preventing the resurgence of COVID-19 or mitigating its impact.

**Keywords:** COVID-19; epidemic; pandemic; human trust; data science; artificial intelligence; deep learning; residential environment; digital health; infoveillance

## Introduction

COVID-19 cases are once again being reported with an increasing trend after laying dormant. To identify what are some lessons learned about preventing the resurgence of COVID-19, Coccia [1] reviewed the literature for factors that could have contributed to the emergence of COVID-19, including environmental factors such as the air quality and population density of a city (see also [2, 3].). At the individual level, however, it remains to explore the role that air quality and residential population density of one's living environment played in the emergence of COVID-19 and its resurgence in HK in late December after a brief period of inactivity (May 22, 2021, to December 23, 2021). Hence, to add to the lessons to be learned from previous phases of the pandemic, we contribute a case study of HK whereby the unique and combined effects of different elements in one's living environment and sociodemographic profiles' effects on the emergence and resurgence of the COVID-19 pandemic were examined using a novel deep learning model whose construction was "anthropomorphized" [4] after Centers for Disease Control and Prevention's (CDC's) socio-ecological system (SES) framework for prevention.

Due to COVID-19's airborne nature and HK's high-density living conditions, the internal built environment of residential buildings has been examined as a potential mechanism for transmission during the pandemic's early phase. This emerging body of research focuses primarily on architectural elements associated with ventilation [5, 6] or conducive to social interaction [7, 8]. The potential role that one's external built environment played in the pandemic's earlier phase has also been examined [9, 10]. It is of note that the environmental factors that put someone at risk for COVID-19 during the pandemic's earlier phase are consistent with the elements in one's internal and external environment that previous studies linked to communicable and non-communicable diseases or general ill health with unspecific causes [11].

In contrast to the emerging, albeit limited, literature on the different aspects of the buildings' built environment put their residents at risk for COVID-19 during the early phase of the pandemic, the role of the built environment during the pandemic's recent resurgence has not been explored, even though the resurgence of COVID-19 was characterized by more infectious variants and more stringent public health and social measures that were constantly evolving to keep up. For example, it remains unknown whether the lengthened homestay was associated with the increase in stringency and the corresponding decrease in mobility since the resurgence of the pandemic has enhanced the effect of the built environment on COVID-19. Neither do we know if the effort to grow a vaccinated population, triggered by the pandemic's resurgence with more contagious variants, may affect the threshold to which the built environment takes effect. Nor do we know which elements in one's built environment are linked to the pandemic's re-emergence and whether the same elements had always been putting individuals at risk during the early waves.

Furthermore, the effects of built environments on COVID-19 infection also exist in the context of the residents' sociodemographic profiles. The social determinants of health are well-studied, including the role of one's living environment [12]. However, while the participants' living environment has been included as one of the social determinants of COVID-19 in at least one study conducted during the pandemic's earlier waves [13], it remains unexplored whether one's sociodemographic profile and built environment may contribute differentially to COVID-19 and their respective contributions may also change over the course of the pandemic. For example, while disadvantaged sociodemographic profiles put individuals at a greater risk of contracting the virus in the community early in the pandemic due to greater exposure to poor workplace hygiene while having limited access to personal protection, it is unclear if the more stringent public health and social measures implemented as a response to the resurgent of the pandemic had enhanced the effects of one's built environment over that of sociodemographic factors due to the restricted mobility of the residential population. Hence, despite separate studies have examined in isolation the effect of a person's sociodemographics, internal environment, or external environment on his/her risk of COVID-19 infection, at least during the earlier phase of the pandemic, their unique and joint contributions to COVID-19 infections had never been explored in the context of their complex interrelationships with one another.

Nevertheless, being able to consider the complex relationships among different aspects of one's environment, including one's internal milieu, household, housing, community, etc., and isolate high-value prevention targets is essential to policy decision and prevention program designs under CDC's social-ecological framework for disease prevention [14, 15]. Notwithstanding the social-ecological framework played a significant role in decision-making concerning policy and prevention programs around the world, data-centric research that aligns with the framework is lacking, which is a challenge if the CDC's social-ecological framework were to be applied to prevent the resurgence of COVID-19. As Coccia [1] observes, well-developed predictive and epidemic models are key to enabling pre-emptive measures against COVID-19, such as improving the effectiveness of public governance. However, while advances in data sciences and analytics (and artificial intelligence, or AI, algorithm more broadly) are making great contributions into areas such as COVID-19's vaccination levels [16, 17], timeliness [18], effectiveness [19], and hesitancy [17]; contact tracing [19], infoveillance with social media [19-21], content analysis [22, 23] identifying misinformation via content analysis [24], identifying misinformation via infodemiology [25, 26], telemedicine [3], they were not designed to represent the effect that the hierarchy of interrelated socio-ecological elements in a person's surrounding has on COVID-19. Hence, if this family of artificial intelligence algorithms were to be deployed in support of the SES framework, these algorithms would lack what Choudhury [27] finds to be essential to the application of artificial intelligence (AI) in clinical settings, i.e. ecological validity – as this family of AI algorithms was not "anthropomorphized" [4] for system think required by patient safety-related decision framework [27] like what the panelists in CDC had exercised when creating the SES framework for prevention.

Here, we "anthropomorphized" an AI algorithm according to the hierarchical organization of different elements in a person's socioecology that had been previously linked to COVID-19 or health more broadly. Specifically, to achieve ecological validity, we created a hierarchical input architecture with features sourced from multiple public databases. An AI algorithm with hierarchically organized and intricately interrelated features was applied to predict 360 buildings' case counts accumulated throughout the pandemic and validate the model with its forecasting performance over rolling 14-day horizons in a two-month period.

## Methods

### Data sources and feature space

The current study examined the built environment and residents' profiles of 360 buildings from public housing estates of three adjacent HK districts whose residents represented 68.69%, 36.17%, and 18.85% of the populations of District A (N=487,200), B (N=424,000), and C (N=316,400), respectively. The three adjacent districts were managed by the same hospital network (called Public Hospital Cluster in HK), whose purpose is to ration operational resources of the hospitals in each cluster so all residents' continuum-of-care needs can be met within the region. The three districts' studied buildings' internal built environment, external built environment, residents' sociodemographic profiles, and the estate-level and tertiary planning units- (the official subregions of a district demarcated for town-planning purposes) level sociodemographic profiles are represented by features extracted from public-domain governmental databases from Census and Statistics Department, Housing Authority, etc., are described in Table 1. The census data were collected from individuals but only available as processed statistics at the levels of buildings, estates, and Tertiary Planning Units (TPUs). Estates are composed of buildings, while both are located within TPUs. The government's Planning Department divided HK into 291 TPUs, demarcated for urban planning purposes. Architectural data were mainly accessed from the Housing Authority of HK. The architectural dataset described the construction types (named after floor plans and constructional details), the year a building was built, the number of floors, etc. Mainly, HK's public housing had standard construction types, only allowing slight variants when a building was constructed in a microenvironment. Thus, the three districts' public residential buildings can be measured with the same features sharing the feature space. We included urban planning data on building-based access to various infrastructures satisfying citizens' diverse living, e.g., distance to market, hospital, parks, and sports stadium. These data were captured from the geographic map in July 2022 from the Lands Department of HK government and measured by driving or walking distance. We accessed COVID-19 spread data from the government's Latest local situation of COVID-19 website, which provided personal details on reported cases, including the date of confirmation and the residential address of the confirmed cases. The data were administered by the HK government's Department of Health. Features in Table 1 were the independent variables for modeling COVID-19 case counts during waves 1–4 (2020-01-23 and 2021-05-21) and wave 5 (2021-12-24 and 2022-07-23). COVID-19 case counts during the study period were extracted from the Department of Health's website and press releases.

Table 1. List of included variables for feature engineering. [a-e]

| Data Source | SES hierarchy | Variable for feature engineering | District A | District B | District C |
|---|---|---|---|---|---|
| | | | | | |
| **Government Open data** | | | | | |
| | **Demographic** | | | | |
| | | Population density (by age)[c] | 16.67% | 17.03% | 19.90% |
| | | Population density (by sex)[c] | 46.70% | 46.16% | 46.43% |
| | | Sex ratio: male / female[a] | 0.909 | 0.885 | 0.89% |
| | | Sex (%, by age group)[b] male and 65+ | 18.98% | 20.65% | 23.53% |
| | | Age (median, by sex)[a] | 44.7 (5.9) | 48.2 (6.0) | 46.5 (5.3) |
| | | Ability to read Chinese (%, by sex)[b,c] | 96.52% | 97.05% | 96.30% |
| | | Ability to write Chinese (%, by sex)[b,c] | 93.98% | 94.90% | 94.12% |
| | | Usual spoken language (#/%, by sex)[a,b,c,d] | 94.13% | 95.78% | 92.58% |
| | | Marital status (#/%, by sex)[a,b,c,d] | 22.99% | 24.32% | 22.01% |
| | | Population by ethnicity (#/%, by sex)[b,c,d] | 48.10% | 47.35% | 46.48% |
| | | Place of birth (%)[a,b]: Mainland China/Macau/Taiwan | 43.54% | 46.26% | 45.56% |
| | | | | | |
| | **Education** | | | | |
| | | Main mode of transport to place of study (%): Mass Transit Railway (MTR)[a] | 20.88% | 34.43% | 18.56% |
| | | Educational attainment (by age group) (#/%): secondary[a,b,c,d] | 51.78% | 50.45% | 50.55% |
| | | Studying population living in the same district (%)[a] | 66.95% | 63.74% | 61.73% |
| | | Place of study (#/%)[b,c,d]: Kowloon | 16.64% | 17.69% | 14.29% |
| | | | | | |
| | **Economic** | | | | |
| | | Working population[a,c] | 824 (369) | 552 (299) | 1063 (260) |
| | | Non-working population[a,c] | 901 (365) | 640 (335) | 1391 (297) |
| | | Working population by | 16.16% | 14.53% | 15.82% |
| | | occupation (#/%)[a,b,c,d] | | | |
| | | Working population by industry (%)[a,b,c] | 29.91% | 32.40% | 23.02% |
| | | Working population by weekly working hours (%)[c] | 6.34% | 6.74% | 6.21% |
| | | Working population by employment status (%)[c] | 91.56% | 89.46% | 91.92% |
| | | Working population by monthly income (%)[c] | 11.58% | 13.46% | 11.51% |
| | | Working population living in the same district (%)[c] | 29.39% | 24.67% | 26.24% |
| | | Employee in working population (%)[a] | 91.94% | 91.75% | 92.29% |
| | | Employer in working population (%)[a] | 1.01% | 1.03% | 1.46% |
| | | Self-employed and unpaid family workers in working population (%)[a] | 6.47% | 7.21% | 5.97% |
| | | Students in non-working population (%)[a] | 25.51% | 25.23% | 26.96% |
| | | Non-students in non-working population (%)[a] | 74.49% | 74.77% | 73.04% |
| | | Monthly income from main employment (median)[a,c] | 19091 (5751) | 15740 (4725) | 13829 (3333) |
| | | Labour force (by sex)[c] | 52.66% | 51.79% | 52.26% |
| | | Labour force participation rate (by sex)[c] | 67.94 | 67.3 | 65.17 |
| | | Weekly usual working hours (median #)[a,b,e] | 44.66 (1.49) | 44.7 (1.48) | 44.63 (1.07) |
| | | Economic activity status (#/%)[b,c,d]: employee | 44.22% | 42.11% | 42.39% |
| | | Place of work (#/%)[b,c,d]: work in the same district of residence | 29.39% | 24.67% | 26.25% |
| | | Main mode of transport to place of work (%)[a]: MTR | 38.76% | 58.67% | 32.36% |
| | | | | | |
| | **Household** | | | | |
| | | Household composition (%)[a,b]: Other households | 16.43% | 25.03% | 16.97% |
| | | Proportion of flat by household size (%)[a,b]: 1-3 household members | 68.37% | 74.61% | 74.47% |
| | | Domestic households by household size[c]: 3 household members | 26.00% | 22.59% | 24.88% |
| | | Domestic households by household composition[c]: Composed of couple and unmarried children | 37.11% | 31.48% | 32.21% |
| | | Domestic households by monthly domestic household income[c]: less than 6,000 | 9.89% | 9.62% | 9.39% |
| | | Domestic household size | 2.9 (0.5) | 2.6 (0.5) | 2.7 (0.5) |
| | | (median and mean)[a,b,c] | | | |
| | | Average (domestic) household size[b] | 2.85 (0.28) | 2.56 (0.37) | 2.70 (0.47) |
| | | Median age of heads of domestic households[a] | 59.6 (4.6) | 59.7 (3.5) | 60.4 (5.9) |
| | | Monthly (domestic) household income (median and #)[a,b,c,e] | 19092 (5751) | 15740 (4725) | 13829 (3333) |
| | | Nationality of heads of domestic households (%)[a]: Chinese | 98.91% | 99.23% | 99.06% |
| | | | | | |
| | **Housing** | | | | |
| | | Monthly domestic household mortgage payment and loan repayment (median)[b,c] | 4113.79 (2211.64) | 2161.76 (507.30) | NA |
| | | Monthly domestic household rent (median)[b,c] | 1727.67 (558.29) | 1710.86 (620.4) | 1553.33 (570.1) |
| | | Monthly domestic household rent to income ratio (median)[c] | 9.84 (1.2) | 13.18 (4.54) | 13.84 (7.47) |
| | | Mortgage payment and loan repayment to income ratio (median)[c] | 17.35 (0.84) | 19.34 (1.87) | 23.43 (2.35) |
| | | Ratio of median rent by income[a,b] | 9.48% | 10.93% | 9.86% |
| | | Occupied quarters by type of housing[c]: Public rental housing | 60.61% | 54.62% | 72.07% |
| | | Domestic households by type of housing[c]: Public rental housing | 60.13% | 55.13% | 72.15% |
| | | Population by type of housing: public rental housing[c] | 58.45% | 52.16% | 70.30% |
| | | Domestic households by number of rooms[c]: three rooms | 44.64% | 33.52% | 37.79% |
| | | Domestic households by tenure of accommodation[c]: Sole tenant | 66.45% | 68% | 81.62% |
| | | Domestic households by monthly domestic household mortgage payment and loan repayment[c] : 20,000HKD and over | 4.98% | 18.97% | 8.13% |
| | | Domestic households by monthly domestic household rent[c] : 10,000HKD and over | 3.99% | 6.73% | 6.56% |
| | | Population by area of residence 5 years ago[c] | 90.49% | 82.15% | 88.99% |
| | | Floor area of accommodation of domestic households (median)[c] | 35.98 (1.56) | 35.86 (5.26) | 34.85 (3.41) |
| | | | | | |
| **Architectural Data** | | | | | |
| | | | | | |
| | | Year of completion | 1991 (10) | 1994 (11) | 1981 (14) |
| | | Age of the building | 31 (10) | 28 (11) | 41 (14) |
| | | Number of corridor | 4 (2) | 3 (2) | 4 (2) |
| | | Number of flat | 611 (240) | 471 (245) | 539 (280) |
| | | Number of floor | 29 (9) | 24 (12) | 22 (11) |
| | | Number of lift | 4 (1) | 4 (2) | 4 (1) |
| | | Indication of re-entrant bay: Yes (%) | 79.89% | 77.78% | 5.13% |
| | | Indication of terrace: Yes (%) | 62.43% | 48.72% | 23.08% |
| | | Number of slides having access to light | 2 (0) | 2 (0) | 1 (1) |
| | | Cardinal orientation (%) | | | |
| | | - East | 9.52% | 12.82% | NA |
| | | - West | 6.88% | 8.55% | NA |
| | | - South | 8.47% | 13.68% | NA |
| | | - North | 7.94% | 5.98% | NA |
| | | - Northeast | 9.52% | 13.68% | NA |
| | | - Northwest | 10.05% | 5.13% | NA |
| | | - Southeast | 5.29% | 15.38% | NA |
| | | - Southwest | 4.76% | 22.22% | NA |
| | | - Southeast & Northwest | 0.53% | 0.00% | NA |
| | | - Northeast & Northwest | 0% | 1% | NA |
| | | - Northeast & Southeast | 0% | 1% | NA |
| | | Proportion of flat by number of non-functional room: 1-2 | 38.38% | 52.06% | 59.33% |
| | | | | | |
| **Urban Planning Data** | | | | | |
| | **External built environment** | | | | |
| | | Distance from building to the closest | | | |
| | | - Hospital (by car) (kilometer) | 2.99 (1.26) | 1.58 (0.84) | 2.28 (0.86) |
| | | - Clinic (by walk) (meter) | 810.17 (418.19) | 472.85 (290.12) | 200.08 (120.21) |
| | | - Market (by walk) (meter) | 659.6 (473.24) | 630.27 (571.42) | 240.85 (122.77) |
| | | - Park (by walk) (meter) | 878.25 (379.05) | 753.97 (370.91) | 313.23 (206.39) |
| | | - Recreational sites (by walk) (meter) | 404.6 (214.58) | 804.27 (284.88) | 213.9 (144.44) |
| | | - Playground or gym (by walk) (meter) | 560.24 (271.3) | 524.43 (259.05) | 651.26 (253.59) |
| | | - Supermarket (by walk) (meter) | 659.02 (497.75) | 267.07 (147.1) | 171.97 (87.4) |
| | | Buildings in the estate (#)[e] | 8 (4) | 9 (5) | 7 (3) |
| | | Recreation sites (≤400m) (#)[e] | 1 (1) | 4 (2) | 2 (1) |
| | | Restaurants and pubs (≤50m) (#)[e] | 0 (1) | 1 (1) | 0 (1) |
| | | Transport stations (≤50m) (#)[e] | 0 (0) | 1 (1) | 0 (0) |

[a]Statistics by Building level.
[b]Statistics by Estate level.
[c]Statistics by Tertiary Planning Unit Group (TPU) level.
[d]#/% representing numbers or prevalence.
[e]# representing numbers.

Before supervising the model prediction and feature selection, COVID-19 case counts extracted were first parameterized as a binary outcome (high- versus low-risk during waves 1-4 and wave 5, respectively), whose cutoff values were selected to maximize the model performance during the training phases of the study periods. In addition to the 10-fold validation performed using COVID-19 case counts extracted from the same period when model-building data were extracted, the wave-5 model was also validated with COVID-19 case counts extracted during the two months (2022-05-22 and 2022-07-23) subsequent to when data were extracted for model development (2021-12-24 and 2022-05-21). An overview of the steps taken to integrate, process, and analyze data, etc., is in Figure 1.

## Developing and validating the model with COVID-19 case counts reported between 2021-12-24 to 2022-05-21

The development of COVID-19 case counts-supervised Multi-Headed Hierarchical Convolutional Neural Networks (MHHCNNs) was based on the studied buildings' internal and external built environments and the residents' sociodemographic profiles. Each input layer was connected to an embedding layer followed by a long short-term memory (LSTM) layer (see Figure 1 for a schematic of its architecture; for the benefit of this hybrid CNN-LSTM approach, please see [28]). While each input layer was comprised of all levels of a single feature, different layers representing features belonging to the same socioecological levels (building-level, estate-level, and TPU-level) were concatenated and served as input to one convolutional layer. We expected that the performance of modeling COVID-19 case counts was enhanced by including elements of buildings' internal and external built environments and the residents' sociodemographic profiles, as all input features were hierarchically structured in alignment with the different levels of the socioecological framework [14]. Performance is measured in terms of the area located under the receiver operating curves (AUCs) and epoch to the MHHCNN model convergence.

MHHCNN also assigned a measure of importance (the Shapely value; see next section for details) to each feature according to its unique contribution to COVID-19 case counts accumulated over the period being modeled. Notably, the unique contribution of each feature is estimated within the context of the contributions of all other features and their interactions. In particular, the hierarchically organized concatenation of the MHHCNN also allows the importance associated with interaction among features from the same or different input layers to be estimated in isolation within the context of other features' contributions, singly or in combination. Figure 2 describes how input, analytics, and outputs were placed together as an AI design flow.

## Quantifying features' unique contribution to COVID-19 case counts to enhance the model's interpretability

Shapely value is a method based on cooperative game theory and has been applied to increase the transparency and interpretability of machine learning models [29]. Shapely value is a numerical expression of the marginal contribution of each parameter to the outcome (i.e. feature importance). The unique contribution of each parameter can be expressed as the degree of change in overall performance when the parameter is excluded. Shapely value is more sensitive, consistent, and accurate than the standardized regression coefficient, which is also aimed at parameterizing unique contributions of individual features but does so in linear models [30]. Notably, SHAP has been dubbed as the key tool in the explainable artificial intelligence's approach to making deep learning models more interpretable to end-users [30]. In the current study, a Shapely value was assigned to each feature value in Table 1 by MHHCNNs. Different Shapely values were assigned to the same features with respect to the COVID-19 case counts modeled during waves 1–4 and wave 5 in Districts A, B, and C. For each model, features were rank-ordered according to their Shapely values, and the rankings of each feature were aggregated across the three districts to provide an overview of the relative importance assigned to top features during waves 1–4 versus wave 5. The Shapely values were obtained by using the Python library *Shap*.

## Validating model (build from data extracted between 2021-12-24 and 2022-05-21) with COVID-19 case counts extracted from 2022-05-22 to 2022-07-23

It is a well-established practice for statistical- or machine-learning models to be cross-validated by dividing the entire dataset into k subsamples, with k-1 subsamples being the model-building set and the remaining one as the model validation set. The random subsampling of model-building and model-validation sets and the subsequent validation of models will be repeated k times. No dependency nor systematic difference is expected between the model-building and model-validation sets. However, dependency and systematic differences between model-building and model-validation sets can be expected if one were to sample the model validation sets in subsequent periods to align with the "intended clinical use" [31] of risk prediction models, which is to predict future risk. Predicting COVID-19 case counts accumulated over time is challenging. On the one hand, the growth of COVID-19 case counts is non-linear and affected by factors that apply to the entire population, such as the virology of the virus, as well as more heterogeneous factors, such as different individuals' socioecology. On the other hand, the growth of COVID-19 is also affected by the different public health and social measures put in place with an intention to curb the spread of the virus at various times in response to the changing case counts. Hence, we have validated our model with data extracted during the two-month period subsequent to when model-building data were extracted with the forward-chaining validation methodology [32].

Figure 3 shows a schematic presentation of the periods from which data were extracted for a forward-chaining validation with three "chains," each consisting of a tandem of training and testing phases. The data extraction period of the first chain's training phase spanned between 2021-12-24 and 2022-05-21, while the testing phase's data extraction period was between 2022-05-22 and 2022-06-04. The training phase of the second chain was based on data extracted over the period that overlapped with both the training and testing phases of the first chain, and the second chain's testing phase was based on data extracted from the 14 days that followed (from 2022-06-05 to 2202-06-18). By the same token, the third chain's training period overlaps with the period when the previous chain's training and testing phases had taken place, with its own 14-day testing phase ending on 2022-07-23. During each testing phase, the performance of forecasting over the 3-, 7-, and 14-day horizons was compared between LSTM deep neural network (DNN) models consisting of both the composite Shapley value of building features and the historical COVID-19 case counts recorded during the training phase and LSTM DNN models that consisted of only historical COVID-19 case counts over the same period. The forward-chaining validation was performed in all three studied districts.

## Ethical considerations

The study was approved by the Secretary Survey and Behavioral Research Ethics Committee of the Chinese University of HK. All data involved in the current study were publicly accessible from corresponding sources. No individual human subjects were involved in the study. All residential buildings and districts involved in the study were anonymized and de-identified to prevent re-identification.

## Results

The study included 345 public residential buildings from Districts A, B, and C. Fifteen buildings were excluded from the analysis because they were designed and used as special housing for seniors or other particular purposes, which made their internal built environment and property management heterogeneous to the general residential buildings. Seventy-five percent of buildings were built before 1999 in District A (143/189), before 2004 in District B (91/117), and before 1999 in District C (30/39). Majority of buildings were high-rise of at least 13 floors (97.4% (184/189), 81.2% (95/117), and 79.5% (31/39) of Districts A, B, and C). The average number of storeys per building was 29 (Standard Deviation (SD)=9.0), 24 (SD=12.5), and 22 (SD=11.4) for the three districts. The median of accumulated cases during COVID-19 waves 1-4 was 1, 1, and 0 for the three districts, respectively. The median of accumulated cases during the COVID-19 wave 5's early spread (between 2021-12-24 and 2022-05-21) was 221 (mean=223, SD=114), 140 (mean=156, SD=102), and 183 (mean=176, SD=96).

### Identifying the optimal cutoff value for modeling COVID-19 case counts

A cutoff value of one COVID-19 case yielded optimal model-training performance of our MHHCNN (AUCs=.90, .90, and .72 for Districts A, B, and C, respectively) when deployed to the case counts of COVID-19 accumulated in studied buildings during waves 1–4 (2020-1-23 to 2021-05-21). In addition, the average 10-fold validations AUCs of our MHHCNN predicting building with at least one COVID-19 case during waves 1-4 were .95, .92, and .87 for Districts A, B, and C, respectively. On the other hand, a cutoff value of 105 COVID-19 cases yielded optimal model-training performance (AUCs=.99, .98, and .95 for Districts A, B, and C, respectively) during the first six months of wave 5 (2021-12-24 to 2022-05-21). The average 10-fold validations AUCs were .91, .94, and .95 for predicting building with at least 105 cases during the first six months of wave 5 for Districts A, B, and C, respectively.

### Quantifying features' unique contribution to COVID-19 case counts with Shapely values

Table 2 reports the top features to which our MHHCNN assigned the greatest importance (parameterized in Shapley value) in predicting COVID-19 case counts during waves 1–4 and the first six months of wave 5 across the three studied districts. Table 2's last two columns represent the sum of inverse-coded ranks across the three districts for waves 1–4 and wave 5, respectively. Notwithstanding the variability across districts, Table 2 shows a consistent pattern of the built environment and sociodemographic features being the primary contributors to COVID-19 case count during wave 5 and waves 1-4, respectively. For example, the number of flats (i.e. the number of distinct households per building or per floor), floors, corridors, and lifts; and the building's cardinal orientation consistently ranked top across the three districts during wave 5. On the other hand, while sociodemographic features showed greater overall scores compared to the overall scores of built environment features during waves 1-4, the ranking of individual sociodemographic features varied across the districts. Nonetheless, even with less consistency compared to the ranking of built environment features during wave 5, the following sociodemographic features were generally ranked high across the districts: The proportion of non-working population, weekly hours of work, monthly income, occupations by industries, and the proportion of the population aged less than 15 years old. With the proportion of the non-working population being the most consistent sociodemographic feature.

Table 2. Each district's ranking of features and their cross-district composite scores during waves 1-4 vs 5.

| | Features' Ranking (District A) | | Features' Ranking (District B) | | Features' Ranking (District C) | | Total Score | |
|---|---|---|---|---|---|---|---|---|
| | Wave 5 | Waves 1-4 | Wave 5 | Waves 1-4 | Wave 5 | Waves 1-4 | Wave 5 | Waves 1-4 |
| | AUC= 0.91 | AUC= 0.95 | AUC= 0.94 | AUC= 0.92 | AUC= 0.95 | AUC= 0.87 | | |
| | | | | | | | | |
| Number of flat | 1 | 2 | 1 | 1 | 1 | | 60 | 39 |
| Maximum number of flat per floor | 7 | | 3 | 11 | 3 | | 50 | 10 |
| Number of floors | 11 | | 4 | | 2 | | 46 | 0 |
| Number of corridors of the building | 2 | | 8 | | 8 | | 45 | 0 |
| Non-working population in building group | 5 | 1 | 10 | 19 | 4 | 20 | 44 | 23 |
| Number of lifts | 3 | | 2 | 4 | 17 | | 41 | 17 |
| Working population in building group | 4 | 20 | 16 | | 14 | | 29 | 1 |
| Cardinal Orientations | 8 | | 7 | | 20 | | 28 | 0 |
| Minimum number of flat per floor | 15 | | 12 | | 11 | | 25 | 0 |
| Population density of the building | 13 | 7 | | | 5 | | 24 | 14 |
| Distance to the secondary closest clinic by walk (meter) | 10 | | 11 | | | 9 | 21 | 12 |
| Number of transport station within 50m radius of the building | | | 5 | 10 | | | 16 | 11 |
| Main mode of transport to place of work: others (private car, company bus/van, taxi or others) | | | | 9 | 6 | | 15 | 12 |
| Proportion of flat with 4 non-functional room | | | 6 | | | 16 | 15 | 5 |
| Median age of heads of | 6 | | | | | | 15 | 0 |
| domestic households | | | | | | | | |
| Place of birth(%): the mainland of China/ Macau/ Taiwan | | | | | 7 | | 14 | 0 |
| Median domestic household size | | | | | 9 | | 12 | 0 |
| Proportion of flat with 3 non-functional room | | | 9 | | | | 12 | 0 |
| Monthly income from main employment: lower quartile | 9 | | | | | | 12 | 0 |
| Occupation (%): craft and related workers, plant and machine operators and assemblers | | 18 | | | 10 | | 11 | 3 |
| Distance to the secondary closest playground or gym by walk (meter) | 20 | | 13 | | | | 9 | 0 |
| Educational attainment (highest | | | | | 12 | | 9 | 0 |
| level attended): secondary | | | | | | | | |
| Weekly usual hours of work of all employment: upper quartile | 12 | | | | | | 9 | 0 |
| Educational attainment (highest level attended): primary and below | | | | | 13 | | 8 | 0 |
| Monthly domestic household income(excluding foreign domestic helpers): lower quartile | 14 | 9 | | | | | 7 | 12 |
| Number of restaurants and pubs within 50m radius of the building | | | 14 | | | | 7 | 0 |
| Weekly usual hours of work of employment (excluding foreign domestic helpers): | | | 15 | 18 | | | 6 | 3 |
| lower quartile | | | | | | | | |
| Proportion of population studying and living in the same district | | | | | 15 | | 6 | 0 |
| Occupation (%): professionals and associate professionals | | 16 | | | 16 | | 5 | 5 |
| Main mode of transport to place of study: school bus | 16 | | | | | | 5 | 0 |
| Industry(%): public administration, education, human health and social work activities, miscellaneous social and personal services | | | 17 | 8 | | | 4 | 13 |
| Distance to the secondary closest market by walk (meter) | 17 | | | | | | 4 | 0 |
| Median age among females | | | | | 18 | | 3 | 0 |
| Occupation (%): craft and related workers (estate level) | 18 | | | | | | 3 | 0 |
| Industry(%): manufacturing, construction, and other industries | | | 18 | | | | 3 | 0 |
| Main mode of transport to place of study: on foot only | | 8 | 19 | | | | 2 | 13 |
| Main mode of transport to place of study: other bus | | | | | 19 | | 2 | 0 |
| Proportion of population aged over 65 | 19 | | | | | | 2 | 0 |
| Indication of lift | | | 20 | | | | 1 | 0 |
| Weekly usual hours of work of all employment: lower quartile | | 3 | | 7 | | | 0 | 32 |
| Monthly income | | | | 5 | | 6 | 0 | 31 |
| from main employment (excluding foreign domestic helpers): upper quartile | | | | | | | | |
| Industry(%): transportation, storage, postal and courier services, information and communication, financing and insurance, real estate, professional and business services | | 11 | | 6 | | | 0 | 25 |
| Proportion of population aged < 15 | | 17 | | | | 1 | 0 | 24 |
| Distance to the closest market by walk (meter) | | | | 15 | | 7 | 0 | 20 |
| Main mode of transport to place of work: MTR | | | | | | 2 | 0 | 19 |
| Usual spoken | | 15 | | | | 8 | 0 | 19 |
| language of aged 5 and over: Putonghua and other Chinese dialects | | | | | | | | |
| Usual spoken language of aged 5 and over: Cantonese | | | | 2 | | | 0 | 19 |
| Monthly income from main employment: upper quartile | | | | | | 3 | 0 | 18 |
| Occupation (%): elementary occupations, skilled agricultural and fishery workers and not classifiable | | | | 3 | | | 0 | 18 |
| Weekly usual working hours of all employment among aged 55-64 | | | | | | 4 | 0 | 17 |
| Occupation (%): managers and administrators | | 4 | | | | | 0 | 17 |
| Indication of re-entrant bay | | | | | | 5 | 0 | 16 |
| Main mode of transport to place of study: bus | | 5 | | | | | 0 | 16 |
| Occupation (%): clerical support, service and sales workers | | 6 | | 20 | | | 0 | 16 |
| Industry(%): import/export, wholesale and retail trades, accommodation and food services | | | | 13 | | 17 | 0 | 12 |
| Weekly usual hours of work of employment (excluding foreign domestic helpers): median | | 14 | | 16 | | | 0 | 12 |
| Distance to the closest clinic by walk (meter) | | | | | | 10 | 0 | 11 |
| Main mode of transport to place of work: bus | | 10 | | | | | 0 | 11 |
| Median monthly | | | | | | 11 | 0 | 10 |
| income from main employment in HKD (estate level) | | | | | | | | |
| Proportion of employer in working population | | 13 | | | | 19 | 0 | 10 |
| Monthly income from main employment: median | | 12 | | | | | 0 | 9 |
| Proportion of non-students in non-working population | | | | | | 12 | 0 | 9 |
| Distance to the secondary closest recreational sites by walk (meter) | | | | 12 | | | 0 | 9 |
| Educational attainment among at any age: secondary (estate level) | | | | | | 13 | 0 | 8 |
| Proportion of flat with household size of 1-3 people | | | | 14 | | | 0 | 7 |
| Monthly domestic | | | | | | 14 | 0 | 7 |
| household income: median | | | | | | | | |
| Educational attainment among aged 15+: secondary (estate level) | | | | | | 15 | 0 | 6 |
| Nationality of heads of domestic households: Chinese | | | | 17 | | | 0 | 4 |
| Distance to the closest hospital by car (kilometer) | | | | | | 18 | 0 | 3 |
| Distance to the closest supermarket by walk (meter) | | 19 | | | | | 0 | 2 |

### Forward-chaining model validations over 3-, 7-, and 14-day forecasting horizons

Figure 4 shows the performance of LSTM DNN models when applied to perform 3-, 7-, and 14-day forward-chaining validations in the three studied districts. The red line in Figure 4 represents the base model's performance that includes only building-level historical COVID-19 case counts from 2021-12-24 to the day before each of the testing phases from the three "chains." On the other hand, the green line in Figure 4 represents the validation model's performance that includes both historical COVID-19 case counts and the composite Shapley value that MHHCNN assigned to each predictive feature used in modeling COVID-19 case counts between 2021-12-24 and 2022-05-21. As the figure shows, the validation model that also included the composite Shapley value has consistently outperformed the base model in terms of the number of epochs required to converge at its performance peak and the magnitude of its performance peak. In particular, forward-chaining validation with a 7-day horizon consistently yielded the best performance; in contrast, forward-chaining validation with a 14-day horizon dropped to the same or below the level of the based model in District C (Figure 4).

## Discussion

The current study shows that different types of factors were differentially linked to COVID-19 case counts during waves 1–4 vs 5 in similar a pattern across the three studied districts, and such variability in predisposing factors during the early phase of the pandemic vs. its resurgence had never been considered by previous studies. For example, during wave 5, COVID-19 case counts were associated with internal built environment features such as the number of distinct households, floors, corridors, and lifts. In addition, the building's cardinal orientation, which is linked to the building's access to light, also ranked among the top-importance features in wave 5. Hence, our findings are consistent with the separate studies that examined in isolation the effects of COVID-19-predisposing built environment features, such as the number of households in the buildings [10] (see also studies on the associated between case counts and multi-family social contact [33, 34] that may be the result of the building having a large number of unique households), floors [9], and corridors [35]. Notably, the COVID-19-predisposing effects of the above-mentioned internal built environment features consistently ranked higher than residential density, which is itself a feature of high importance in our study and the literature [36, 37]. On the other hand, while sociodemographic features showed greater overall scores during waves 1-4 compared to that of built environment features, the ranking of individual sociodemographic features varied across the districts. Nonetheless, the following sociodemographic features consistently ranked top in terms of their importance to predicting COVID-19 case counts: The proportion of non-working population, weekly hours of work, monthly income, occupations by industries, and the proportion of the population aged <15. Our finding that a higher proportion of the non-working population is associated with a higher COVID-19 case count is consistent with a recent study on the linkage between sociodemographic profile and urban health [38], while the association between COVID-19 risk and the remaining sociodemographic features such as weekly hours of work, monthly income [39], occupations by industries, and the proportion of the population aged <15 [36], were consistent with the result of a local contact survey [40].

Taken together, these findings are consistent with the evidenced association between sociodemographics and COVID-19 and between the built environment and COVID-19. Nevertheless, the sociodemographic and built environment features modeled in the current study had never been examined together in a single study nor had they been examined in the context of other built environment factors and sociodemographic factors. Nor had their effectiveness been investigated during early phrases versus epidemic resurgence.

Our finding that the internal built environment played a bigger role since wave 5 than before is consistent with the fact that Omicron, the COVID-19 variant that dominated wave 5, is more infectious and has begotten more cases in the community than previous waves. Hence, while wave 5 had triggered more stringent public health and social measures that decreased the mobility and lengthened homestay of the population, it also increased the likelihood that those who needed to go out into the community be returned infectious and to a stationary residential population congregated in closed space. Consequently, the residential built environment features that can enhance the likelihood of different households in the same building coming into contact with one another were therefore assigned greater importance than sociodemographic factors in wave 5 to identify buildings with a cluster of 105 cases or more, even though the sociodemographic profile of the building was of greater importance during waves 1-4 in classifying which buildings would have gotten its first case. It is also consistent with our findings that, with a larger number of residents having been vaccinated, more infected cases are needed to enable enough viral load to accumulate in the building for its built environment elements, e.g., those connectivity elements that induce social mixing, to take effect.

Most notably, the current study has demonstrated that time-invariant environmental and sociodemographic factors could add significant value to the rolling historical case counts in forecasting subsequent ones over a two-week horizon across different periods and geographical areas. The impact of environmental and sociodemographic factors on the evolution of COVID-19 highlighted the need to put urban planning and public housing policies at the heart of public health and social policies for curtailing the re-emergence of COVID-19 or the emergence of the next pandemic. Another implication of our findings is methodological. Forecasting in epidemiology is mostly performed using compartmental models [41], which have served as the basis of many public health and social policies for curbing COVID-19 [42]. However, compartmental models were designed for epidemiological forecasts at the national or tertiary level. Therefore, they are incompatible with forecasting at the sub-regional level with high variability [43]. Meanwhile, statistical-, machine-, and deep-learning models have been applied to forecast COVID-19 at the population and local levels but with mixed results [44]. Furthermore, no published COVID-19 forecasting model has considered the effect of time-invariant factors concerning one's built environment. Hence, our approach to forward chaining over multiple horizons added value to the forecasting research in epidemiology.

More generally, AI algorithms, such as deep learning ones, have been criticized as being a black box that produces accurate but unexplainable results. Consequently, it's been challenging for AI algorithms to be trusted in clinical settings as both the algorithm's accuracy and explainability are critical to an AI algorithm being trusted by its clinical end-users [45]. Hence, we anthropomorphized our AI algorithm according to CDC's SES framework using an MHHCNN architecture, whose objective is to represent those hierarchically interrelated elements (i.e. potential prevention targets) in the residents' socioecology as layered and differentially dependent input features. Not only have our efforts to achieve ecological validity in constructing our algorithm yielded highly accurate models, but they have also enabled the quantification of the unique and combined contributions of each feature and every level of the features in the pool to the accuracy of model prediction and forecasting. Our effort to maximize the explainability of our AI algorithm is consistent with the literature on the "trustworthiness" of AI, advocating for algorithmic transparency that makes visible the reasoning behind machine learning decisions to enable users' control and address their concerns of accountability [45] (see [46, 47] for a discussion on algorithmic transparency at the technical levels). In addition, having algorithmic transparency that inspires the trustworthiness of end-users is essential for interprofessional collaborations in clinical settings [48] and multi-stakeholder participation involving different sectors [45]. In our study, the multi-stakeholder participation enabled by the transparency of the algorithm is further enhanced by its architecture, which integrates multiple siloed databases sourced from different governmental agencies - resulting in findings that are not only explainable but also actionable.

## Limitations

This study has a few limitations. Aside from accessing only three public databases and examining only a few buildings, the current study is limited by our exclusive focus on public housing. Even though in 2022, 45% of the HK population was living in public housing (rental or subsidized home ownership), our study's exclusive focus on public housing limits the applicability of our findings and systematically excludes potentially key contributors in public housing residents' network of COVID-19 transmission as residents from public and private housing are not completely insulated from one another. For example, in HK, privately owned housing residents and nearby public housing residents share many of the same external environment, since HK's private and public housing are often built in close proximity. In addition, some internal built environment that has been linked to ventilation, such as re-entrant bays (or air shaft), can be found in both public and private housings. Hence, further studies can be designed to expand the research scope by including both public and private residential buildings. In addition, the models reported here did not include any epidemiological characteristics of the various COVID-19 variants dominant during different study periods, nor have our models taken into consideration the timing and the effect of the public health and social measures that evolved with the pandemic.

## Conclusions

Consistently across the three studied districts, the internal built environment of public housing residents is a stronger predisposing factor for COVID-19 infection than the residents' sociodemographics during the resurgence of the pandemic in HK, and the relative importance among the environmental and sociodemographic factors significantly forecasted COVID-19 case count over a 7-day horizon during the resurgence period. On the other hand, while sociodemographic factors were more dominant during the earlier phase of the pandemic, which specific sociodemographic factors were of the greatest importance to COVID-19 infection differed for different districts. As our socioecologically anthropomorphized deep learning algorithms have demonstrated that residents' internal and external environment and sociodemographics may have subjected them to differential risk of COVID-19 infection, there is an implication that vaccination campaigns and public health and social measures against COVID-19 can be geographically targeted, with different housing estates or residential areas being assigned different priority. These findings shall apply to regions with high-rise residential buildings of dense population.

## Acknowledgements

The research is supported by the Strategic Public Policy Research Funding Scheme (project number S2019.A4.015.19S). K.O.K. thanks the Health and Medical Research Fund (reference no. INF-CUHK-1, 17160302, 18170312, CID-CUHK-A, COVID1903008), the General Research Fund (reference nos. 14112818, 24104920), Wellcome Trust Fund (UK, 200861/Z/16/Z), and Group Research Scheme of The Chinese University of Hong Kong.

## Data Availability

The data sets analyzed during this study are publicly available. The census data of the residential population's sociodemographics are available from the Internet source of the Census and Statistics Department of the HK government. The architectural data of public housing buildings are accessible from the Housing Authority of HK. The COVID-19 infection data are available from the Department of Health of the HK government. The data on distance measures to different facilities can be obtained from the Lands Department of the HK government.

## Conflicts of Interest

none declared.

## Abbreviations

AI: Artificial intelligence
AUC: Area located under the receiver operating curve
CNN: Convolutional neural network
COVID-19: Coronavirus Disease 2019
DNN: Deep neural networks
LSTM: Long short-term memory
MTR: Mass Transit Railway
SD: standard deviation
SES: Socio-ecological system
TPU: Tertiary planning unit

## References

1. Coccia M. Sources, diffusion and prediction in COVID-19 pandemic: lessons learned to face next health emergency. AIMS Public Health. 2023;10(1):145-68. doi: 10.3934/publichealth.2023012.
2. Núñez-Delgado A, Bontempi E, Coccia M, Kumar M, Farkas K, Domingo JL. SARS-CoV-2 and other pathogenic microorganisms in the environment.

Environmental Research. 2021 2021/10/01/;201:111606. doi: https://doi.org/10.1016/j.envres.2021.111606.
3. Bontempi E, Coccia M. International trade as critical parameter of COVID-19 spread that outclasses demographic, economic, environmental, and pollution factors. Environmental Research. 2021 2021/10/01/;201:111514. doi: https://doi.org/10.1016/j.envres.2021.111514.
4. Glikson E, Woolley AW. Human Trust in Artificial Intelligence: Review of Empirical Research. Academy of Management Annals. 2020;14(2):627-60. doi: 10.5465/annals.2018.0057.
5. Bhagat RK, Davies Wykes MS, Dalziel SB, Linden PF. Effects of ventilation on the indoor spread of COVID-19. Journal of Fluid Mechanics. 2020;903:F1. doi: 10.1017/jfm.2020.720.
6. Dai H, Zhao B. Association of the infection probability of COVID-19 with ventilation rates in confined spaces. Building Simulation. 2020 2020/12/01;13(6):1321-7. doi: 10.1007/s12273-020-0703-5.
7. Kembel SW, Meadow JF, O'Connor TK, Mhuireach G, Northcutt D, Kline J, et al. Architectural Design Drives the Biogeography of Indoor Bacterial Communities. PLOS ONE. 2014;9(1):e87093. doi: 10.1371/journal.pone.0087093.
8. Sun C, Zhai Z. The efficacy of social distance and ventilation effectiveness in preventing COVID-19 transmission. Sustainable Cities and Society. 2020 2020/11/01/;62:102390. doi: https://doi.org/10.1016/j.scs.2020.102390.
9. Kan Z, Kwan M-P, Wong MS, Huang J, Liu D. Identifying the space-time patterns of COVID-19 risk and their associations with different built environment features in Hong Kong. Science of The Total Environment. 2021 2021/06/10/;772:145379. doi: https://doi.org/10.1016/j.scitotenv.2021.145379.
10. von Seidlein L, Alabaster G, Deen J, Knudsen J. Crowding has consequences: Prevention and management of COVID-19 in informal urban settlements. Building and Environment. 2021 2021/01/15/;188:107472. doi: https://doi.org/10.1016/j.buildenv.2020.107472.
11. Azuma K, Yanagi U, Kagi N, Kim H, Ogata M, Hayashi M. Environmental factors involved in SARS-CoV-2 transmission: effect and role of indoor environmental quality in the strategy for COVID-19 infection control. Environmental Health and Preventive Medicine. 2020 2020/11/03;25(1):66. doi: 10.1186/s12199-020-00904-2.
12. Marmot M. Social determinants of health inequalities. The Lancet. 2005;365(9464):1099-104. doi: 10.1016/S0140-6736(05)71146-6.
13. Morante-García W, Zapata-Boluda RM, García-González J, Campuzano-Cuadrado P, Calvillo C, Alarcón-Rodríguez R. Influence of Social Determinants of Health on COVID-19 Infection in Socially Vulnerable Groups. International Journal of Environmental Research and Public Health. 2022;19(3):1294. PMID: doi:10.3390/ijerph19031294.
14. Krug EG, Mercy JA, Dahlberg LL, Zwi AB. The world report on violence and health. Lancet. 2002 Oct 5;360(9339):1083-8. PMID: 12384003. doi: 10.1016/s0140-6736(02)11133-0.
15. The Social-Ecological Model: A framework for prevention. Centers for Disease Control and Prevention and Health Resources and Services Administration

(CDC); 2022; Available from:
https://www.cdc.gov/violenceprevention/about/social-ecologicalmodel.html.
16. Coccia M. Optimal levels of vaccination to reduce COVID-19 infected individuals and deaths: A global analysis. Environmental Research. 2022 2022/03/01/;204:112314. doi: https://doi.org/10.1016/j.envres.2021.112314.
17. Coccia M. Improving preparedness for next pandemics: Max level of COVID-19 vaccinations without social impositions to design effective health policy and avoid flawed democracies. Environmental Research. 2022 2022/10/01/;213:113566. doi: https://doi.org/10.1016/j.envres.2022.113566.
18. Benati I, Coccia M. Global analysis of timely COVID-19 vaccinations: improving governance to reinforce response policies for pandemic crises. International Journal of Health Governance. 2022;27(3):240-53. doi: 10.1108/IJHG-07-2021-0072.
19. Benati I, Coccia M. Effective Contact Tracing System Minimizes COVID-19 Related Infections and Deaths: Policy Lessons to Reduce the Impact of Future Pandemic Diseases. 2022. 2022 2022-08-24;12(3):15. doi: 10.5296/jpag.v12i3.19834.
20. Xu Q, Nali MC, McMann T, Godinez H, Li J, He Y, et al. Unsupervised Machine Learning to Detect and Characterize Barriers to Pre-exposure Prophylaxis Therapy: Multiplatform Social Media Study. JMIR Infodemiology. 2022 2022/4/28;2(1):e35446. doi: 10.2196/35446.
21. Chandrasekaran R, Desai R, Shah H, Kumar V, Moustakas E. Examining Public Sentiments and Attitudes Toward COVID-19 Vaccination: Infoveillance Study Using Twitter Posts. JMIR Infodemiology. 2022 2022/4/15;2(1):e33909. doi: 10.2196/33909.
22. Chen S, Yin SJ, Guo Y, Ge Y, Janies D, Dulin M, et al. Content and sentiment surveillance (CSI): A critical component for modeling modern epidemics. Frontiers in Public Health. 2023 2023-March-16;11. doi: 10.3389/fpubh.2023.1111661.
23. Chin H, Lima G, Shin M, Zhunis A, Cha C, Choi J, et al. User-Chatbot Conversations During the COVID-19 Pandemic: Study Based on Topic Modeling and Sentiment Analysis. J Med Internet Res. 2023 2023/1/27;25:e40922. doi: 10.2196/40922.
24. Al-Rawi A, Fakida A, Grounds K. Investigation of COVID-19 Misinformation in Arabic on Twitter: Content Analysis. JMIR Infodemiology. 2022 2022/7/26;2(2):e37007. doi: 10.2196/37007.
25. Kyabaggu R, Marshall D, Ebuwei P, Ikenyei U. Health Literacy, Equity, and Communication in the COVID-19 Era of Misinformation: Emergence of Health Information Professionals in Infodemic Management. JMIR Infodemiology. 2022 2022/4/28;2(1):e35014. doi: 10.2196/35014.
26. Galbraith E, Li J, Rio-Vilas VJD, Convertino M. In.To. COVID-19 socio-epidemiological co-causality. Scientific Reports. 2022 2022/04/06;12(1):5831. doi: 10.1038/s41598-022-09656-1.
27. Choudhury A. Toward an Ecologically Valid Conceptual Framework for the Use of Artificial Intelligence in Clinical Settings: Need for Systems Thinking, Accountability, Decision-making, Trust, and Patient Safety Considerations in Safeguarding the Technology and Clinicians. JMIR Hum Factors. 2022 Jun 21;9(2):e35421. PMID: 35727615. doi: 10.2196/35421.
28. Song J, Zhang L, Xue G, Ma Y, Gao S, Jiang Q. Predicting hourly heating load in a district heating system based on a hybrid CNN-LSTM model. Energy and Buildings. 2021 2021/07/15/;243:110998. doi: https://doi.org/10.1016/j.enbuild.2021.110998.
29. Lipovetsky S, Conklin M. Analysis of regression in game theory approach. Applied Stochastic Models in Business and Industry. 2001 2001/10/01;17(4):319-30. doi: https://doi.org/10.1002/asmb.446.
30. Chung WJ, Liu C. Analysis of input parameters for deep learning-based load prediction for office buildings in different climate zones using eXplainable Artificial Intelligence. Energy and Buildings. 2022 2022/12/01/;276:112521. doi: https://doi.org/10.1016/j.enbuild.2022.112521.
31. Goldstein BA, Navar AM, Pencina MJ, Ioannidis JPA. Opportunities and challenges in developing risk prediction models with electronic health records data: a systematic review. Journal of the American Medical Informatics Association. 2017;24(1):198-208. doi: 10.1093/jamia/ocw042.
32. Tashman LJ. Out-of-sample tests of forecasting accuracy: an analysis and review. International Journal of Forecasting. 2000 2000/10/01/;16(4):437-50. doi: https://doi.org/10.1016/S0169-2070(00)00065-0.
33. Nash D, Qasmieh S, Robertson M, Rane M, Zimba R, Kulkarni SG, et al. Household factors and the risk of severe COVID-like illness early in the U.S. pandemic. PLOS ONE. 2022;17(7):e0271786. doi: 10.1371/journal.pone.0271786.
34. Kwok KO, Wei WI, Huang Y, Kam KM, Chan EYY, Riley S, et al. Evolving Epidemiological Characteristics of COVID-19 in Hong Kong From January to August 2020: Retrospective Study. J Med Internet Res. 2021 Apr 16;23(4):e26645. PMID: 33750740. doi: 10.2196/26645.
35. Bartolucci A, Templeton A, Bernardini G. How distant? An experimental analysis of students' COVID-19 exposure and physical distancing in university buildings. International Journal of Disaster Risk Reduction. 2022 2022/02/15/;70:102752. doi: https://doi.org/10.1016/j.ijdrr.2021.102752.
36. Guan C, Tan J, Hall B, Liu C, Li Y, Cai Z. The Effect of the Built Environment on the COVID-19 Pandemic at the Initial Stage: A County-Level Study of the USA. Sustainability. 2022;14(6):3417. PMID: doi:10.3390/su14063417.
37. Yip TL, Huang Y, Liang C. Built environment and the metropolitan pandemic: Analysis of the COVID-19 spread in Hong Kong. Building and Environment. 2021 2021/01/15/;188:107471. doi: https://doi.org/10.1016/j.buildenv.2020.107471.
38. Larsen FB, Pedersen MH, Friis K, Glümer C, Lasgaard M. A Latent Class Analysis of Multimorbidity and the Relationship to Socio-Demographic Factors and Health-Related Quality of Life. A National Population-Based Study of 162,283 Danish Adults. PLOS ONE. 2017;12(1):e0169426. doi: 10.1371/journal.pone.0169426.
39. Oh TK, Choi J-W, Song I-A. Socioeconomic disparity and the risk of contracting COVID-19 in South Korea: an NHIS-COVID-19 database cohort study. BMC Public Health. 2021 2021/01/15;21(1):144. doi: 10.1186/s12889-021-10207-y.

40. Kwok KO, Cowling B, Wei V, Riley S, Read JM. Temporal variation of human encounters and the number of locations in which they occur: a longitudinal study of Hong Kong residents. J R Soc Interface. 2018 Jan;15(138). PMID: 29367241. doi: 10.1098/rsif.2017.0838.
41. Chowell G. Fitting dynamic models to epidemic outbreaks with quantified uncertainty: A primer for parameter uncertainty, identifiability, and forecasts. Infectious Disease Modelling. 2017 2017/08/01/;2(3):379-98. doi: https://doi.org/10.1016/j.idm.2017.08.001.
42. Bakhta A, Boiveau T, Maday Y, Mula O. Epidemiological Forecasting with Model Reduction of Compartmental Models. Application to the COVID-19 Pandemic. Biology. 2021;10(1):22. PMID: doi:10.3390/biology10010022.
43. Kong L, Duan M, Shi J, Hong J, Chang Z, Zhang Z. Compartmental structures used in modeling COVID-19: a scoping review. Infectious Diseases of Poverty. 2022 2022/06/21;11(1):72. doi: 10.1186/s40249-022-01001-y.
44. Chandra R, Jain A, Singh Chauhan D. Deep learning via LSTM models for COVID-19 infection forecasting in India. PLOS ONE. 2022;17(1):e0262708. doi: 10.1371/journal.pone.0262708.
45. Schoenherr JR, Abbas R, Michael K, Rivas P, Anderson TD. Designing AI Using a Human-Centered Approach: Explainability and Accuracy Toward Trustworthiness. IEEE Transactions on Technology and Society. 2023;4(1):9-23. doi: 10.1109/TTS.2023.3257627.
46. Busuioc M. Accountable Artificial Intelligence: Holding Algorithms to Account. Public Administration Review. 2021;81(5):825-36. doi: https://doi.org/10.1111/puar.13293.
47. Hapke H, Nelson C. Building machine learning pipelines: O'Reilly Media; 2020. ISBN: 1492053163.
48. Payne P, Lele O, Johnson B, Holve E. Enabling Open Science for Health Research: Collaborative Informatics Environment for Learning on Health Outcomes (CIELO). J Med Internet Res. 2017 2017/07/31;19(7):e276. doi: 10.2196/jmir.6937.

## Figures

Figure 1

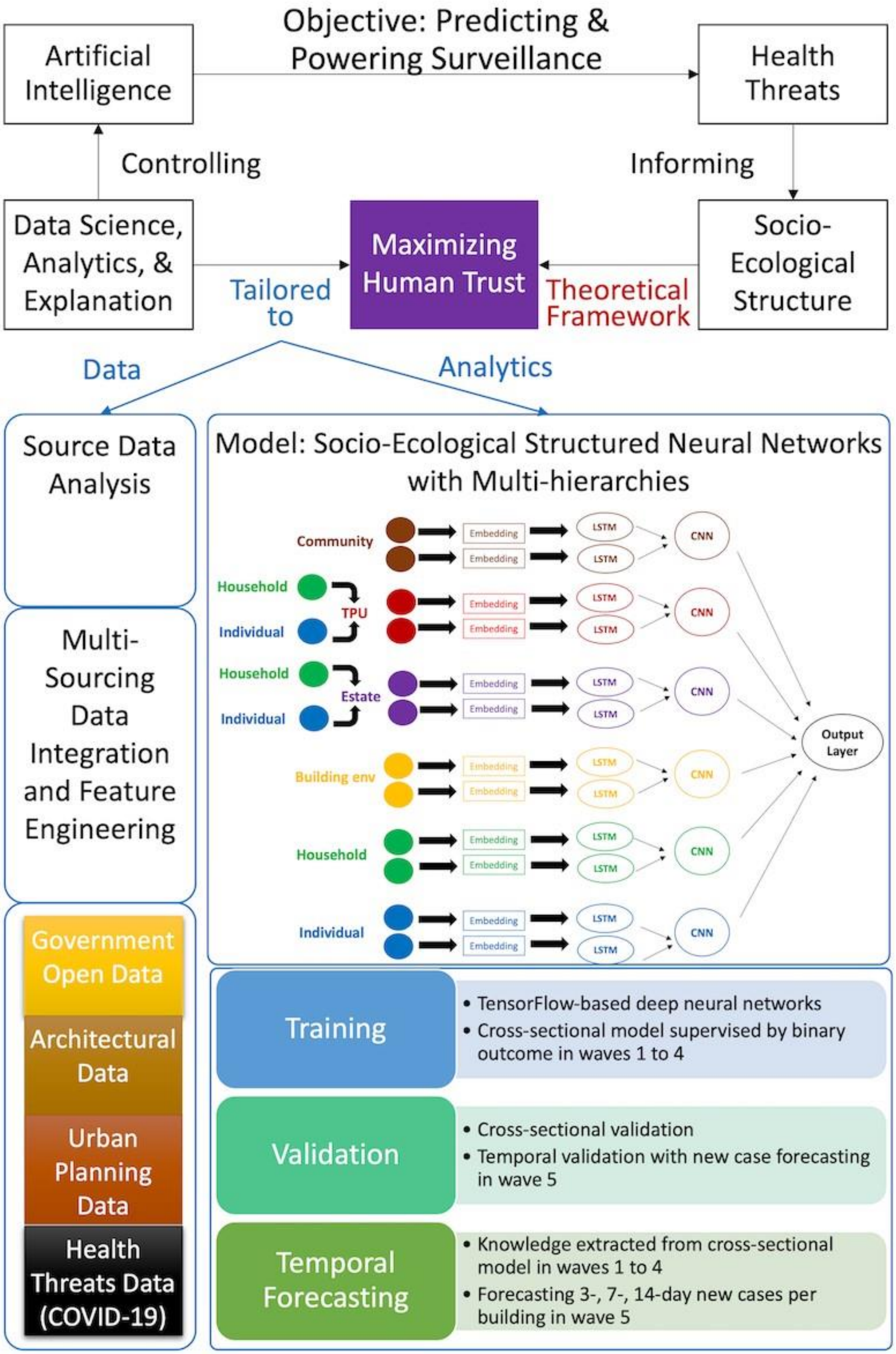

Figure 2

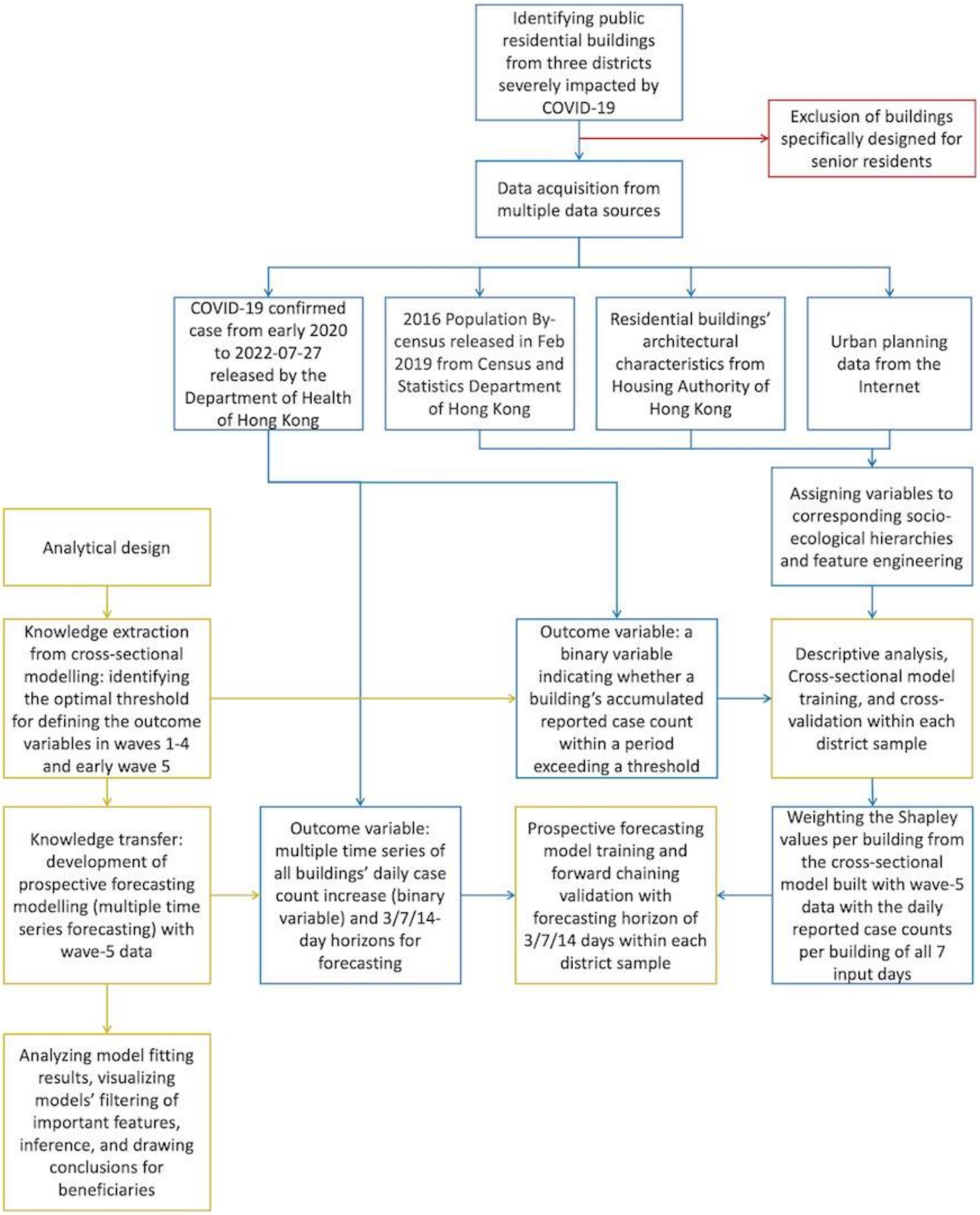

Identifying public residential buildings from three districts severely impacted by COVID-19
Exclusion of buildings specifically designed for senior residents
Data acquisition from multiple data sources
COVID-19 confirmed case from early 2020 to 2022-07-27 released by the Department of Health of Hong Kong
2016 Population By-census released in Feb 2019 from Census and Statistics Department of Hong Kong
Residential buildings' architectural characteristics from Housing Authority of Hong Kong
Urban planning data from the Internet
Assigning variables to corresponding socio-ecological hierarchies and feature engineering
Analytical design
Knowledge extraction from cross-sectional modelling: identifying the optimal threshold for defining the outcome variables in waves 1-4 and early wave 5
Outcome variable: a binary variable indicating whether a building's accumulated reported case count within a period exceeding a threshold
Descriptive analysis, Cross-sectional model training, and cross-validation within each district sample
Knowledge transfer: development of prospective forecasting modelling (multiple time series forecasting) with wave-5 data
Outcome variable: multiple time series of all buildings' daily case count increase (binary variable) and 3/7/14-day horizons for forecasting
Prospective forecasting model training and forward chaining validation with forecasting horizon of 3/7/14 days within each district sample
Weighting the Shapley values per building from the cross-sectional model built with wave-5 data with the daily reported case counts per building of all 7 input days
Analyzing model fitting results, visualizing models' filtering of important features, inference, and drawing conclusions for beneficiaries

Figure 3

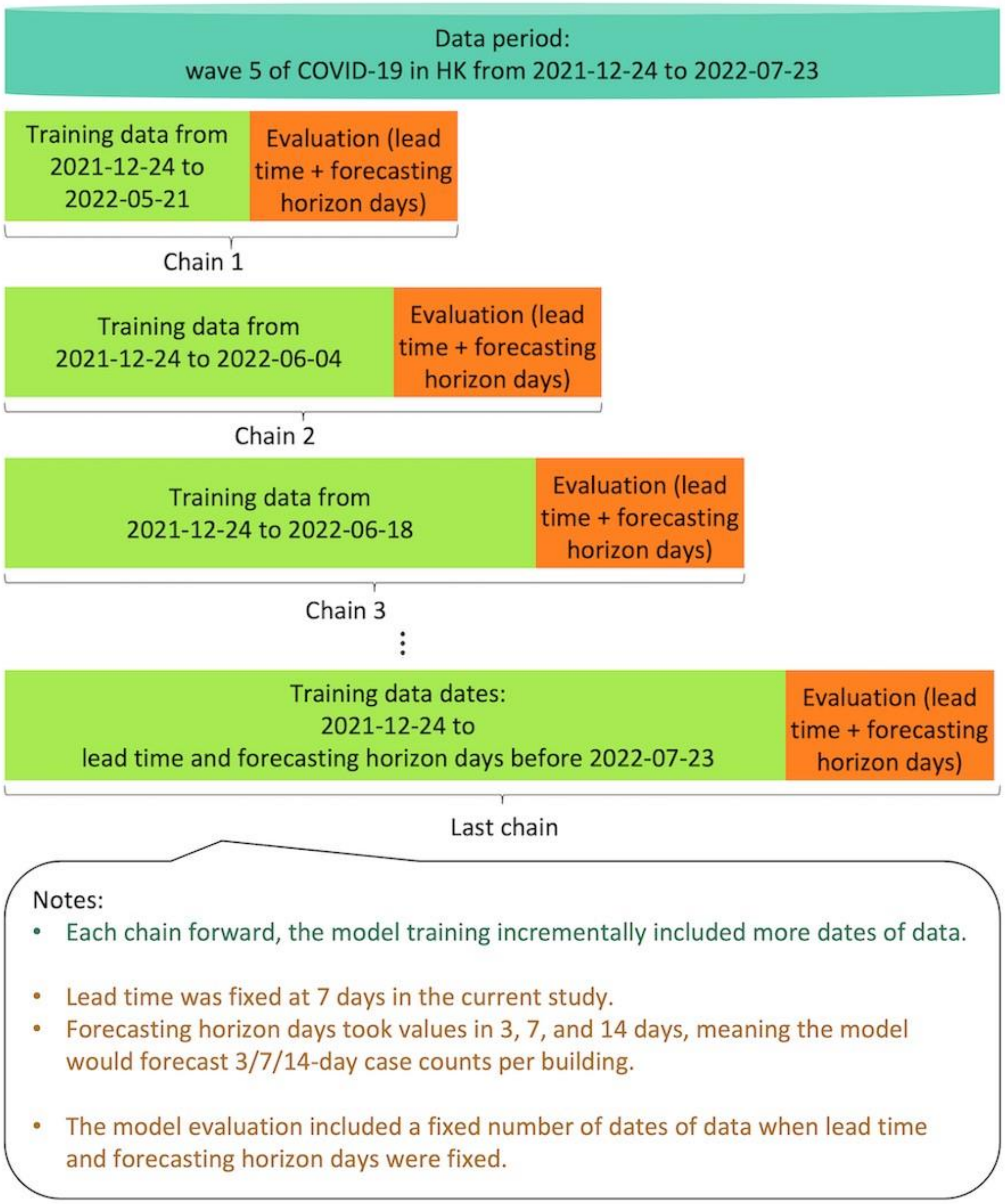
Data period:
wave 5 of COVID-19 in HK from 2021-12-24 to 2022-07-23
Training data from 2021-12-24 to 2022-05-21
Evaluation (lead time + forecasting horizon days)
Chain 1
Training data from 2021-12-24 to 2022-06-04
Evaluation (lead time + forecasting horizon days)
Chain 2
Training data from 2021-12-24 to 2022-06-18
Evaluation (lead time + forecasting horizon days)
Chain 3
Training data dates: 2021-12-24 to lead time and forecasting horizon days before 2022-07-23
Evaluation (lead time + forecasting horizon days)
Last chain
Notes:
• Each chain forward, the model training incrementally included more dates of data.
• Lead time was fixed at 7 days in the current study.
• Forecasting horizon days took values in 3, 7, and 14 days, meaning the model would forecast 3/7/14-day case counts per building.
• The model evaluation included a fixed number of dates of data when lead time and forecasting horizon days were fixed.

Figure 4

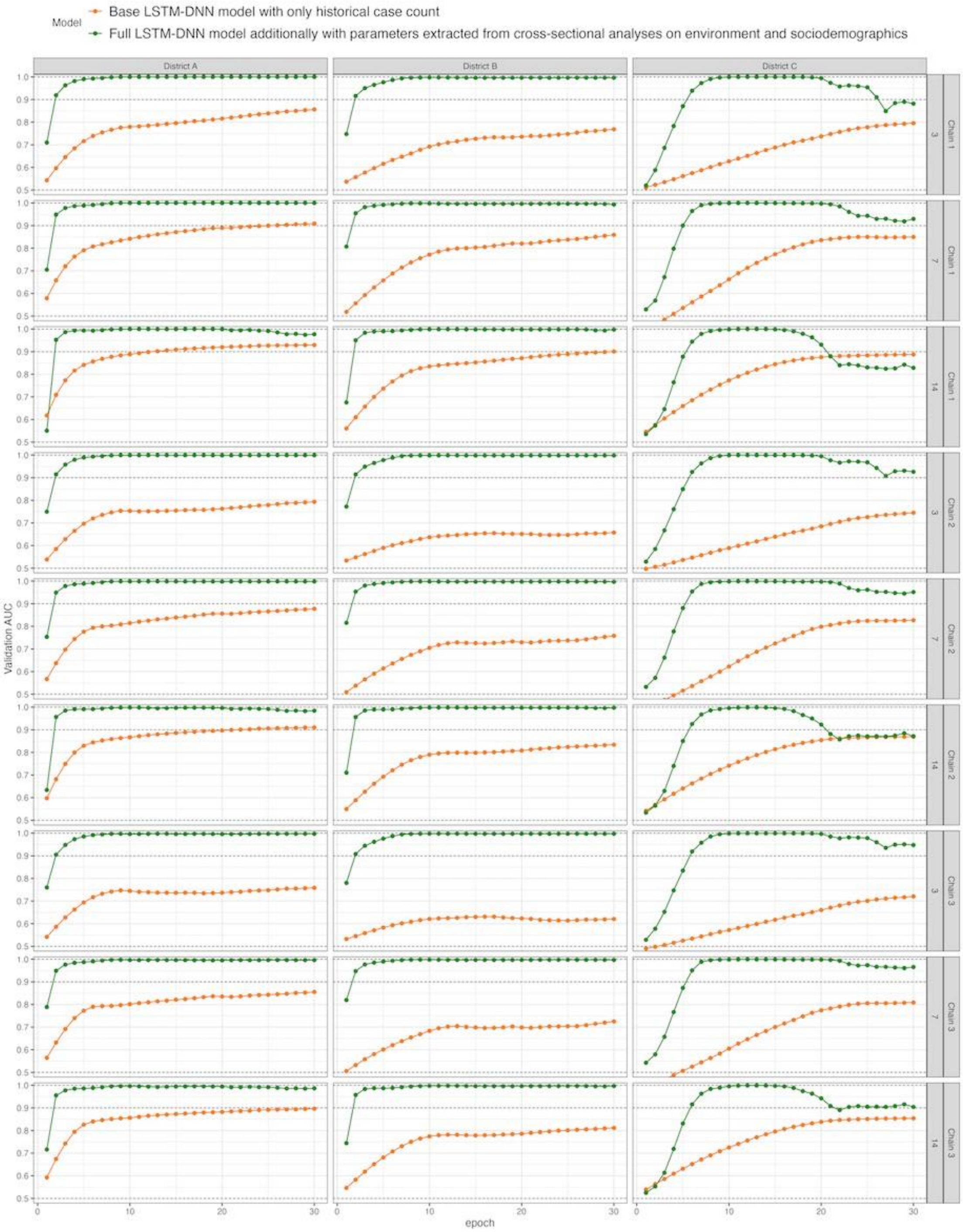
Model
Base LSTM-DNN model with only historical case count
Full LSTM-DNN model additionally with parameters extracted from cross-sectional analyses on environment and sociodemographics
District A
District B
District C
Chain 1
Chain 2
Chain 3
Validation AUC
epoch